\documentclass[%
 reprint,
nofootinbib,
 amsmath,amssymb,
 aps,
pre
]{revtex4-2}

\usepackage{graphicx}
\usepackage{dcolumn}
\usepackage{bm}
\usepackage[english]{babel}
\usepackage{upgreek}
\usepackage{subcaption}

\newcommand{\mc}[1]{\mathcal{#1}}

\newcommand{\half}[0]{\frac{1}{2}}

\newcommand{\kn}[0]{\mathcal{A}}

\newcommand{\uT}{[\text{eV}]}
\newcommand{\un}{[\text{cm}^{-3}]}
\newcommand{\ur}{[\text{cm}]}
\newcommand{\urate}{[\text{ps}^{-1}]}
\newcommand{\uaden}{[\text{g/cm}^{2}]}
\newcommand{\adenig}{(\rho R)_\mathrm{ign}} 
\newcommand{\Tign}{T_\mathrm{ign}}

\begin{document}

\preprint{APS/123-QED}

\title{Improved Ion Heating in Fast Ignition by Pulse Shaping}

\author{Henry Fetsch}
 \email{hfetsch@princeton.edu}
\author{Nathaniel J. Fisch}%
\affiliation{%
 Department of Astrophysical Sciences\\
 Princeton University, Princeton, NJ 08540
}%

\date{\today}

\begin{abstract}

The fast ignition paradigm for inertial fusion offers increased gain and tolerance of asymmetry by compressing fuel at low entropy and then quickly igniting a small region. Because this hotspot rapidly disassembles, the ions must be heated to ignition temperature as quickly as possible, but most ignitor designs directly heat electrons. A constant-power ignitor pulse, which is generally assumed, is suboptimal for coupling energy from electrons to ions. Using a simple model of a hotspot in isochoric plasma, a novel pulse shape to maximize ion heating is presented in analytical form. Bounds are derived on the maximum ion temperature attainable by electron heating only. Moreover, arranging for faster ion heating allows a smaller hotspot, improving fusion gain. Under representative conditions, the optimized pulse can reduce ignition energy by over $20\%$.

\end{abstract}

\maketitle


\section{Introduction}
\label{sec_intro}

In conventional approaches to inertial confinement fusion (ICF) \cite{Edwards_Patel_Lindl_Atherton_Glenzer_Haan_Kilkenny_Landen_Moses_Nikroo_et,Atzeni_2009}, an external driver compresses the fuel through a sequence of carefully-timed shocks, which are usually designed to approximate adiabatic compression \cite{Kidder_1979}. The resulting configuration is isobaric, meaning that pressure is approximately uniform throughout the fuel interior. This equilibrium takes the form of a central hotspot with high temperature and low density surrounded by colder and denser fuel. If the hotspot temperature and areal density are high enough, the hotspot ignites, initiating a thermonuclear burn that propagates into the surrounding fuel \cite{Tabak_Hinkel_Atzeni_Campbell_Tanaka_2006, Craxton_Anderson_Boehly_Goncharov_Harding_Knauer_McCrory_McKenty_Meyerhofer_Myatt_et}. While ICF has seen major advancements \cite{ICF_Collab,Zylstra_Hurricane_Callahan_Kritcher_Ralph_Robey_Ross_Young_Baker_Casey_et}, including reaching ignition \cite{Christopherson_Hurricane_Weber_Kritcher_Nora_Salmonson_Tran_Milovich_Maclaren_Hinkel_et}, outstanding challenges remain ahead of practical inertial fusion energy. Instabilities during compression are a major problem because they can induce an asymmetric implosion, introduce impurities, and reduce burn efficiency \cite{Edwards_Patel_Lindl_Atherton_Glenzer_Haan_Kilkenny_Landen_Moses_Nikroo_et,Craxton_Anderson_Boehly_Goncharov_Harding_Knauer_McCrory_McKenty_Meyerhofer_Myatt_et}.

Fast ignition (FI) schemes mitigate this problem. In FI, compression is designed to be isochoric, meaning that density is roughly uniform, and formation of a hotspot is avoided; ideally, temperature is uniform and well below ignition temperatures \cite{Tabak_Hammer_Glinsky_Kruer_Wilks_Woodworth_Campbell_Perry_Mason_1994}. Isochoric compression involves lower acceleration during the implosion and so is more robust to capsule and drive asymmetries. Additionally, the compressed fuel can reach higher density for the same driver energy \cite{Atzeni_1995, Tabak_Hinkel_Atzeni_Campbell_Tanaka_2006,Atzeni_Meyer-Ter-Vehn_2004, Clark_Tabak_2007,Kodama_Shiraga_Shigemori_Toyama_Fujioka_Azechi_Fujita_Habara_Hall_Izawa_et}. However, the compressed fuel will not spontaneously ignite; it requires a `spark' from some external energy source to ignite a small hotspot region.

This ignitor can deliver energy to the hotspot in various ways, including electrons \cite{Tabak_Hammer_Glinsky_Kruer_Wilks_Woodworth_Campbell_Perry_Mason_1994,Honrubia_Meyer-ter-Vehn_2008,Malkin_Fisch_2002,Kodama_Shiraga_Shigemori_Toyama_Fujioka_Azechi_Fujita_Habara_Hall_Izawa_et}, protons \cite{Fernandez_Albright_Beg_Foord_Hegelich_Honrubia_Roth_Stephens_Yin_2014}, heavy ions \cite{Fernandez_Albright_Beg_Foord_Hegelich_Honrubia_Roth_Stephens_Yin_2014,Hegelich_Jung_Albright_Fernandez_Gautier_Huang_Kwan_Letzring_Palaniyappan_Shah_et,Regan_Schlegel_Tikhonchuk_Honrubia_Feugeas_Nicolai_2011}, and soft x-rays \cite{Lee_Robinson_Pasley_2020,Hu_Goncharov_Skupsky_2012}. A universal feature is that the timescale $t_p$ of heating should be shorter than the hydrodynamic timescale $t_c$ of hotspot disassembly \cite{Atzeni_Meyer-Ter-Vehn_2004,Tabak_Hinkel_Atzeni_Campbell_Tanaka_2006}. Most ignitor designs involve fast particle beams, which deposit energy primarily into electrons in the hotspot. The need to transfer energy from electrons to ions leads to the requirement that $t_p$ is longer than the timescale $t_\nu$ of interspecies collisional energy transfer. Heating electrons too quickly is counterproductive because the collision rate becomes smaller for hotter electrons and because excess electron pressure works to accelerate hydrodynamic expansion.

The full requirement that $t_\nu < t_p < t_c$ has been noted for example by \cite{Tabak_Hinkel_Atzeni_Campbell_Tanaka_2006}, but in FI literature the lower bound on $t_p$ is discussed less widely than the need to outpace hotspot disassembly. This is understandable because the upper bound already places technically challenging demands on the ignitor output power \cite{Zuegel_Borneis_Barty_Legarrec_Danson_Miyanaga_Rambo_Leblanc_Kessler_Schmid_et,Tabak_Hammer_Glinsky_Kruer_Wilks_Woodworth_Campbell_Perry_Mason_1994}. Nonetheless, by analogy to the careful shaping of the driver pulse used for implosions, it is reasonable to suppose that ignitor performance is sensitive to the temporal shape of the pulse. We therefore ask: what ignitor pulse shape is most effective at heating ions?

We address this problem using a simple model of an expanding hotspot in isochoric plasma before ignition. Most studies assume a flat pulse of width $t_p$, but we show (Fig.~\ref{fig_maxT_tp}) that this leads to suboptimal ion heating. We calculate the pulse shape (Eq.~\ref{eq_Pi_opt_ti_x} and Eq.~\ref{eq_Pi_opt_t}) that maximizes the ion temperature that can be reached at given areal density, a critical ignition parameter. We also derive a bound (Eq.~\ref{eq_Ti_hat}) on the ion temperature that can be reached before the plasma ignites. We find that optimized pulse shaping can lead to a more than $20\%$ reduction in the energy requirement for the fast ignitor pulse. Our results are independent of the ignitor design except for the assumption that energy is deposited initially into electrons. These findings could aid in designing efficient FI heating schemes.

This paper is organized as follows. In \S\ref{sec_model} we define the hotspot model and derive evolution equations for radius and temperature. In \S\ref{sec_heating} we find conditions for optimal ion heating, first for an instructive example and then for an expanding hotspot. In \S\ref{sec_params} we assign numerical values to the constants of the model. In \S\ref{sec_numerical_solutions} we compare the optimized pulse to numerical solutions for other pulse shapes. In \S\ref{sec_ignition_conditions} we demonstrate how changes in pulse shape affect the ability of a hotspot to ignite. Finally, in \S\ref{sec_discussion} we discuss extensions to this model and their potential influence on FI development.

\section{Model Setup}
\label{sec_model}

\subsection{Assumptions}
\label{sec_assumptions}

Here we define a reduced model of a fast ignition hotspot that will allow us to derive analytical expressions for pulse shape and temperature bounds. As background cold fuel, we take isochoric plasma of average ion mass $\bar m_i$, ionization state $Z=1$, electron density $n_0$, and mass density $\rho_0 = \bar m_i n_0$. We begin at a time ${t=0}$ at which a small amount of hotspot heating has already occurred so that the thermal and degeneracy pressures of the cold fuel can be taken to be much smaller than the hotspot pressure. The hotspot is modeled as a sphere of plasma with ion and electron temperatures $T_i$ and $T_e$ respectively, radius $r$, and electron density $n$.

As initial conditions in the hotspot, we take starting temperatures $T_i = T_0$ and $T_e = 3T_0$,\footnote{The reasons for starting at slightly elevated electron temperature are rather technical but not important for the main results. The assumption that the hotspot expands at speed $u_s$ requires that the hotspot pressure is much greater than the cold fuel pressure. If the starting temperatures are too low then loss terms, particularly for the electrons, can cause the model to lose-self-consistency; in the extreme case, $T_e$ becomes negative. Additionally, starting near the optimal electron temperature (cf. \S\ref{sec_optimal_Te}) makes direct comparison easier between naive and optimized models. Except for cases where the model breaks self-consistency, changes to this starting temperature have a negligible impact because the final temperatures are much higher than $T_0$.} starting radius $R = R_0$, and starting density $n = n_0$ that is the same as in the cold fuel. Using these initial conditions, we define normalized variables
\begin{align}
    x = R/R_0, && \theta_e = T_e/T_0, && \theta_i = T_i/T_0,
\end{align}
and note that $n = x^{-3} n_0$.

To expose the key effect, we model the hotspot as simply as possible. The hotspot is assumed to have uniform temperature and density, and to be spherically symmetric. Ignitor power is assumed to be deposited uniformly in the hotspot, and we ignore changes to power deposition caused by changing temperature, density, and radius. The actual spatial distribution of ignitor power depends on ignitor design. Our homogeneity assumption allows us to isolate temporal pulse-shaping effects from spatial effects. 

Hotspot expansion is more deleterious in isochric FI than in conventional isobaric ICF because of the larger pressure differential \cite{Atzeni_Meyer-Ter-Vehn_2004,Tabak_Hinkel_Atzeni_Campbell_Tanaka_2006,Atzeni_1995}. The hotspot edge is taken to expand symmetrically and the speed is estimated as that of the material behind a strong shock ${u_s = \sqrt{3 p/ 4\rho_0}}$ where $p = nT_e + nT_i$ is the hotspot pressure. The strong shock assumption is common in fast ignition studies because the hotspot pressure is much greater than the ambient pressure \cite{Atzeni_Meyer-Ter-Vehn_2004,Piriz_Sanchez_1998,Tabak_Hinkel_Atzeni_Campbell_Tanaka_2006}. The interior is taken to expand uniformly. We note that acoustic waves and heat diffusion in the hotspot go some way toward equilibrating the interior and extending the validity of the uniformity assumption.

The rates $\nu_{ee}$ and $\nu_{ii}$ for electrons and ions each to become maxwellian are taken to be faster than all other processes of interest, and the mean free path $L$ is assumed to be much less than $R$ even for suprathermal particles, so the hotspot can be treated as a two-temperature fluid. 

Under these assumptions, we proceed to quantify the dominant power loss terms to due work, radiation, conduction, and interspecies collisions. The neglect of alpha particle power deposition at this stage is discussed further in \S\ref{sec_discussion}.

\subsection{Temperature Evolution}
\label{sec_power_balance}

The time evolution of hotspot parameters is governed by
\begin{equation}
\label{eq_power_balance}
\begin{split}
    C_V \dot T_e &= - W_e - P_{ie} - P_r - P_c + P_\ell ,
    \\
    C_V \dot T_i &= -W_i + P_{ie} ,
    \\
    \dot R &= u_s ,
\end{split}
\end{equation}
where $C_V$ is the isochoric heat capacity of ideal gas, which is the same for each species. It will be helpful to work in normalized variables, so we begin by defining an expansion rate constant $\sigma$ such that
\begin{align}
\label{eq_x_dot}
    u_s/r_0 = \half \sigma x^{-3/2} \sqrt{\theta_e + \theta_i} .
\end{align}

The rate of work $W_s$ for species $s$ is $W_s = p_s 4\pi R^2 u_s$ where $p_s$ is the partial pressure of species $s$. In normalized variables, this takes the form
\begin{equation}
\label{eq_Ws_defn}
\begin{split}
    W_s/C_VT_0 = \sigma x^{-5/2} \theta_s \sqrt{\theta_e + \theta_i} .
\end{split}
\end{equation}

The power $P_{ie}$ transferred from electrons to ions can be written in terms of the rate $\nu$ as
\begin{align}
\label{eq_Pie_defn}
    P_{ie}/C_V T_0 = \nu x^{-3} \frac{\theta_e - \theta_i}{\theta_e^{3/2}}
\end{align}
when $m_e/\bar m_i \ll T_e/T_i$, where $m_e$ is the electron mass. The Coulomb logarithm appearing in $\nu$ is taken to be constant.

For conditions of interest, the power $P_r$ lost to radiation is dominated by Bremsstrahlung, given in terms of the rate $\beta$ as
\begin{align}
\label{eq_Pr_defn}
    P_r/C_VT_0 = \beta x^{-3} \theta_e^{1/2}
\end{align}
and the hotspot is optically thin to Bremsstrahlung.

The power $P_c$ lost to thermal conduction is taken to be dominated by electron losses \cite{Atzeni_Meyer-Ter-Vehn_2004,Freidberg_2007}.  In terms of the rate $\kappa$, the heat loss to conduction is
\begin{align}
\label{eq_Pc_defn}
    P_c/C_V T_0 = \kappa x \theta_e^{7/2} .
\end{align}

Finally, the normalized ignitor power $\Pi_\ell = P_\ell/C_V T_0$ is the quantity that we are interested to control in order to maximize $\theta_i$. The total pulse energy is $E_p$ and we define the pulse timescale $t_p$ such that $\textbf{max}[P_\ell(t)] = E_p/t_p$.

Formulas for $\sigma, ~\nu, ~\beta, ~\text{and}~ \kappa$ are derived in \S\ref{sec_params}. Each of these rates depends on the initial scales $T_0, ~r_0, ~n_0$.

\subsection{Radial Dependence}

The time evolution in Eq.~\ref{eq_power_balance} is a highly nonlinear system of coupled differential equations in three variables. Noting that $x(t)$ is monotonic, we can reduce the system to two equations by defining $\theta_e^\prime = d\theta_e/dx$ and $\theta_i^\prime = d\theta_i/dx$ and changing variables by the relation ${\dot x = u_s/r_0}$. The new system of equations is
\begin{align}
\label{eq_theta_e_prime}
    \theta_e^\prime &= -2\frac{\theta_e}{x} - \frac{2\nu}{\sigma} \frac{(\theta_e - \theta_i)}{x^{3/2}\theta_e^{3/2} (\theta_e + \theta_i)^{1/2}} + \Phi(x) ,
    \\
\label{eq_theta_i_prime}
    \theta_i^\prime &= -2\frac{\theta_i}{x} + \frac{2\nu}{\sigma} \frac{(\theta_e - \theta_i)}{x^{3/2}\theta_e^{3/2} (\theta_e + \theta_i)^{1/2}} ,
\end{align}
where we have defined  
\begin{equation}
\label{eq_Phi_defn}
    \Phi(x) \doteq \frac{2x^{3/2}}{\sigma\sqrt{\theta_e + \theta_i}}\left(\Pi_\ell - \beta x^{-3} \theta_e^{1/2} - \kappa x \theta_e^{7/2}\right)
\end{equation}
to be the net non-mechanical energy into the hotspot per unit increase in radius. Surprisingly, we will find these coupled differential equations to be solvable for certain $\Phi(x)$, including the one that drives optimal heating.

\section{Heating Problem}
\label{sec_heating}

\subsection{Optimal Electron Temperature}
\label{sec_optimal_Te}

The problem that we address here can be simply posed as follows. What value of $T_e$ causes $T_i$ to increase as rapidly (in time or in radius) as possible? Given a desired profile of $T_e$, ignitor power $P_\ell$ can readily be chosen to achieve it. It is instructive to compare to a simpler problem. Given a fixed-volume box, described by the same parameters as the hotspot except holding $u_s = 0$, ion temperature evolution is given by
\begin{align}
\label{eq_fixedV_theta_i_dot}
   \textbf{(fixed V)} ~~ \dot \theta_i = \nu \frac{\theta_e - \theta_i}{\theta_e^{3/2}} .
\end{align}

To extremize $\dot \theta_i$ with respect to $\theta_e$, we solve ${(d\dot\theta_i / d \theta_e) = 0}$, finding ${\theta_e = 3\theta_i}$, which is a maximum. The numerator ${(\theta_e - \theta_i)}$ means that heat flows more quickly when the temperature separation is large, an expected property of heat transport. The denominator $\theta_e^{3/2}$ captures the peculiar feature of plasmas that collisions become rarer at high temperature. These competing effects create an optimal electron temperature at three times the ion temperature. Therefore, if electron heating overshoots the optimal temperature, it will take more time to heat the ions; by contrast, if the species collided as a neutral gas, it would be advantageous to heat the electrons as quickly as possible.

We proceed to the problem of ion heating in an expanding sphere. The dynamics are now further complicated by energy lost to mechanical work during expansion. Because expansion rate increases with electron temperature, we can expect the optimal $\theta_e$ to be lower for a given $\theta_i$ than in the fixed-volume case. Finding $\dot\theta_i$ from Eq.~\ref{eq_power_balance} and maximizing it with respect to $\theta_e$ yields an algebraic equation without a closed-form solution. However, maximizing $\theta_i^\prime$ yields a tractable solution.

Beyond simplifying the analysis, $\theta_i^\prime$ is actually a more important target for optimization. The constraint ${t_p < t_c}$ appeared because the ignitor must heat the hotspot before it cools by doing mechanical work while expanding into the cold fuel. Additionally, the areal density is an important parameter for ignition and scales as $\rho R \sim x^{-2}$. Thus it is of interest to maximize the increase in ion temperature per increment of radius.

Starting with Eq.~\ref{eq_theta_i_prime} and evaluating ${(d\theta_i^\prime/d\theta_e)=0}$ yields a quadatric equation in $\theta_e$ with positive solution
\begin{align}
\label{eq_theta_e_opt}
    \theta_e = \frac{3 + \sqrt{33}}{4} \theta_i .
\end{align}
The $-$ branch is not only a minimum of $\theta_i^\prime$ but also entails negative electron temperature, so we have discarded it. We define a constant ${c_0 \doteq (3 + \sqrt{33})/4 \approx 2.19}$, and observe that the optimal electron temperature ${\theta_e = c_0\theta_i}$ is indeed less than the $3\theta_i$ optimum of the fixed volume case.

\subsection{Optimal Pulse Shape}

Using the prescription for $\theta_e$ in Eq.~\ref{eq_theta_e_opt}, we can solve for a net energy input per radius $\Phi(x)$ that achieves the needed $\theta_e$, which in turn yields the optimal ignitor power $\Pi_\ell(t)$. We define the sum of temperatures $\zeta = \theta_e + \theta_i$ and obtain its evolution equation by adding Eq.~\ref{eq_theta_e_prime} and Eq.~\ref{eq_theta_i_prime} to find $\zeta^\prime = - 2\zeta/x + \Phi$. This equation can be separated to obtain
\begin{align}
\label{eq_x2et_prime}
    \frac{d}{dx}\left(x^2\zeta\right) = x^2 \Phi .
\end{align}

We note that $d(x^2\zeta)/dx = (c_0 + 1)d(x^2\theta_i)/dx$. Now from Eq.~\ref{eq_theta_i_prime} along with Eq.~\ref{eq_theta_e_opt}, we find that $\theta_i$ evolves according to
\begin{align}
\label{eq_x2theta_i_prime}
    \frac{d}{dx}\left(x^2\theta_i\right) = \frac{2\nu c_1}{\sigma} \frac{x^{1/2}}{\theta_i}
\end{align}
where ${c_1 \doteq (c_0-1)/c_0^{3/2}(c_0+1)^{1/2} \approx 0.21}$. Thus, using Eq.~\ref{eq_x2et_prime}, the optimal ignitor power as a function of $\theta_i$ and of $x$ is
\begin{equation}
\label{eq_Pi_opt_ti_x}
\Pi_\mathrm{opt} = \frac{\nu c_1 \sqrt{c_0 + 1}}{x^3 \theta_i^{1/2}} + \frac{\beta c_0^{1/2} \theta_i^{1/2}}{x^3} + \kappa x c_0^{7/2} \theta_i^{7/2}
\end{equation}
when $\theta_e = c_0\theta_i$. If ignitor power is limited by some $\Pi_\mathrm{max} > \Pi_\mathrm{opt}$ then the delivered power should be $\Pi_\ell = \Pi_\mathrm{max}$ until $\Pi_\mathrm{opt}$ drops below $\Pi_\mathrm{max}$. Similarly, if the initial conditions have $\theta_e < c_0\theta_i$, the ignitor should deliver power $\Pi_\mathrm{max}$ until $\theta_e$ reaches $c_0\theta_i$, then continue at $\Pi_\mathrm{opt}$. If initially ${\theta_e > c_0\theta_i}$, power should be held at ${\Pi_\ell = 0}$ until $\theta_e = c_0\theta_i$.

If ignitor power is unlimited, we can find a closed-form expression for optimal $\Pi_\ell(t)$. Solving Eq.~\ref{eq_x2theta_i_prime} for $\theta_i(x)$ gives
\begin{align}
\label{eq_theta_i_x}
    \theta_i(x) = \frac{1}{x^2} \sqrt{1 + \frac{8 \nu c_1}{7\sigma}\left(x^{7/2} - 1\right)} ,
\end{align}
allowing Eq.~\ref{eq_Pi_opt_ti_x} to be written as a function of $x$ only. At this point $\Pi_\mathrm{opt}$ can be written explicitly as a function of time by solving Eq.~\ref{eq_x_dot}; the calculation is presented in Appendix~\ref{app_temporal_shape}.

\section{Plasma Parameters}
\label{sec_params}

In this section we outline formulas for the rate coefficients introduced in \S\ref{sec_power_balance}. Everywhere in this work, Boltzmann's constant $k_B$ is absorbed into the temperature so that $T_e$ and $T_i$ have units of energy (eV).

The expansion rate $\sigma$ is defined to satisfy ${R_0 (\sigma/2) x^{-3/2} \sqrt{\theta_e + \theta_i}=\sqrt{3 (T_e + T_i)n/ 4\bar m_i n_0 }}$. For fully ionized equimolar DT plasma with Coulomb logarithm $\lambda$, the collision rate $\nu$ satisfies \cite{Hazeltine_2004}
\begin{equation}
    T_0 \nu \frac{\theta_e - \theta_i}{x^3\theta_e^{3/2}} = \left(\frac{1}{2m_D} + \frac{1}{2m_T}\right) \frac{m_e^{1/2} e^4 n \lambda}{3\sqrt{2}\pi^{3/2}T_e^{3/2}} (T_e - T_i) .
\end{equation}

We take $C_V$ to be the heat capacity of ideal gas with $\gamma = 5/3$, so if the hotspot contains $N$ electrons, then ${C_V = 3N/2 = 2\pi R^3 n}$. The Bremsstrahlung rate $\beta$ satisfies \cite{Freidberg_2007}
\begin{equation}   
    C_V T_0 \beta x^{-3}\theta_e^{1/2} = \frac{2^{7/2} \pi^{1/2} e^6}{3 c^3 h m_e^{3/2}} n^2 \frac{4\pi R^3}{3} T_e^{1/2} .
\end{equation}

Classical heat flux is nominally infinite across the hotspot-cold fuel discontinuity, so a gradient lengthscale less than the radius must be chosen. Heat flux is also suppressed in regions with strong gradients \cite{Atzeni_Meyer-Ter-Vehn_2004}. We take the lengthscale to be $R$ but incorporate an unknown dimensionless constant $f_c$ in calculating the thermal diffusivity. The rate $\kappa$ satisfies \cite{Krommes_2018,Atzeni_1995}
\begin{equation}
    C_V T_0 \kappa x \theta_e^{7/2} = 4\pi R^2 f_c\frac{T_e}{R_0} \frac{0.957 T_e^{5/2}}{\lambda e^4 m_e^{1/2}}
\end{equation}
and we will estimate $f_c = 0.1$.

The coefficients are given by the following formulas:
\begin{align}
    \sigma\urate &= 1.52\times 10^{-6} R_0\ur^{-1} T_0\uT^{1/2}
    \\
    \nu\urate &= 1.32\times 10^{-21} \lambda n_0\un T_0\uT^{-3/2}
    \\
    \beta\urate &= 7.04\times 10^{-26} n_0\un T_0\uT^{-1/2}
    \\
    \kappa\urate &= 3.87\times 10^{9} \lambda^{-1} n_0\un^{-1} R_0\ur^{-2} T_0\uT^{5/2}
\end{align}

We take the Coulomb logarithm as constant ${\lambda = 8}$, a representative value in the regimes of interest; corrections are generally small. For hotspot density ${\rho = \bar m_i n}$, the areal density is ${\rho R\uaden = 4.18\times 10^{-24} x^{-2} n_0\un R_0\ur}$ and, as defined in Eq.~\ref{eq_kn_defn}, ${\kn = 1.83\times 10^{-16} \lambda n_0\un R_0\ur^{-1} T_0\uT^{-2}}$

\section{Numerical Solutions}
\label{sec_numerical_solutions}

The system of Eq.~\ref{eq_theta_e_prime} and Eq.~\ref{eq_theta_i_prime} is not generally solvable analytically for arbitrary pulse shapes. We numerically integrated the system for three trial pulse shapes other than the optimized pulse Eq.~\ref{eq_Pi_opt_ti_x}. The pulse shapes are labeled as flat, rising sawtooth, and falling sawtooth, defined in terms of $t_p$ as
\begin{equation}
\label{eq_pulse_forms}
\begin{split}   
    P_\mathrm{flat}(t;t_p) =& E_p/t_p ~ : ~ 0 < t < t_p ,
    \\
    P_\mathrm{rise}(t;t_p) =& t E_p/2t_p ~ : ~ 0 < t < 2t_p ,
    \\
    P_\mathrm{fall}(t;t_p) =& E_p (1 - t/2t_p) ~ : ~ 0 < t < 2t_p ,
\end{split}
\end{equation}
where, for all pulses, power is zero outside of the specified time interval. Note that for each of the pulse shapes in Eq.~\ref{eq_pulse_forms}, the peak power is ${P_\mathrm{max} = E_p/t_p}$. For the optimized pulse, $t_p$ sets the power limit ${\Pi_\mathrm{max} = P_\mathrm{max}/C_VT_0}$ for fixed $E_p$.  Because this limit forces the pulse shape to differ from the analytical form in Eq.~\ref{eq_Pi_opt_t}, we numerically integrated the optimized pulse as defined in Eq.~\ref{eq_Pi_opt_ti_x}.

It is worth noting that, for all pulse shapes, $P_\ell$ represents the power ultimately deposited in electrons by the ignitor. The specific ignitor design, for example using a short-pulse laser to accelerate electrons from a foil into the hotspot, determines how $P_\ell$ is related to the actual laser power. In this work, we specify only $P_\ell$ so that our results are independent of the details of the ignitor.

\begin{figure}
\begin{subfigure}{0.48\columnwidth}
    \includegraphics[width=\columnwidth]{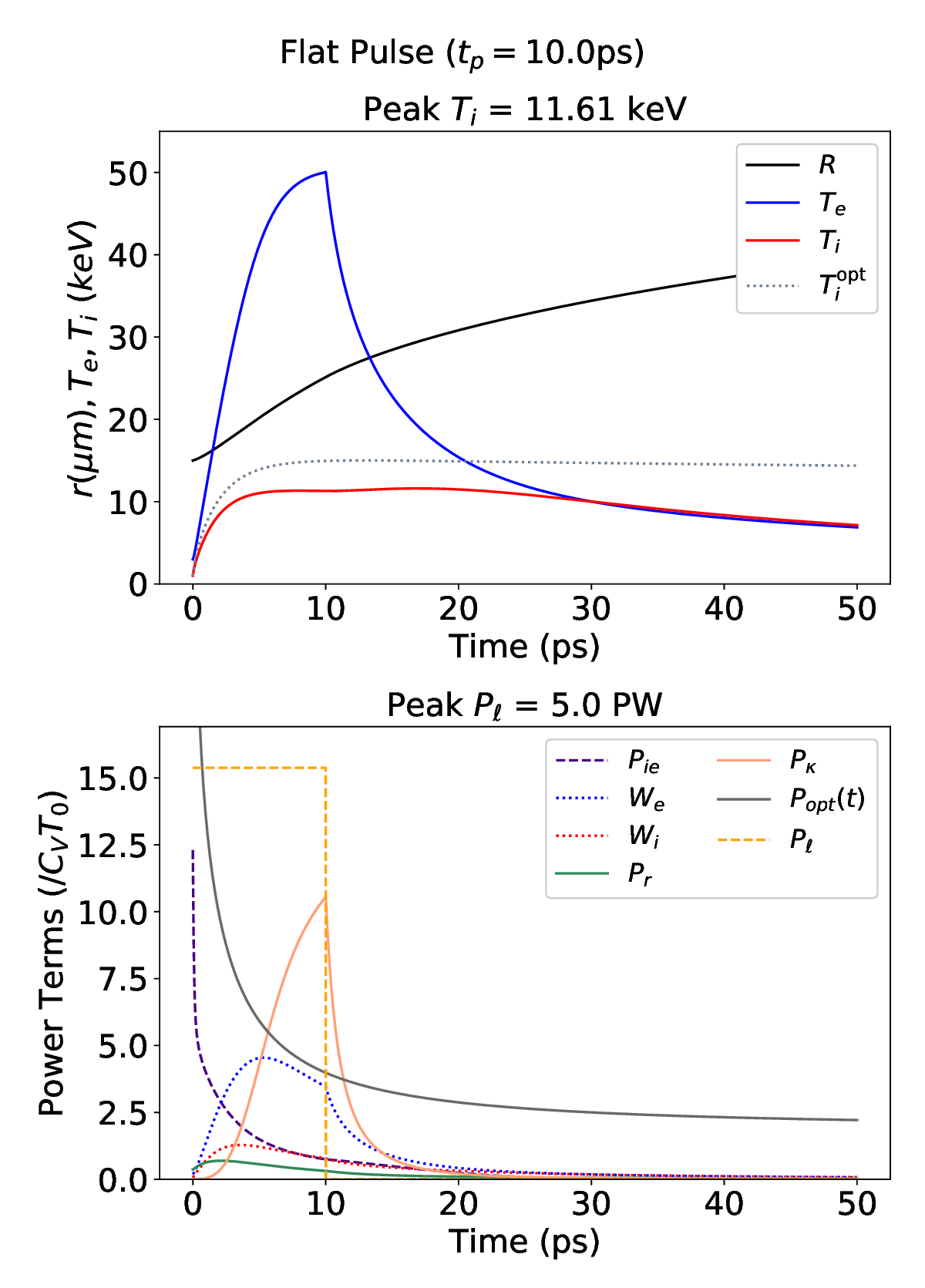}
\end{subfigure}
\begin{subfigure}{0.48\columnwidth}
    \includegraphics[width=\columnwidth]{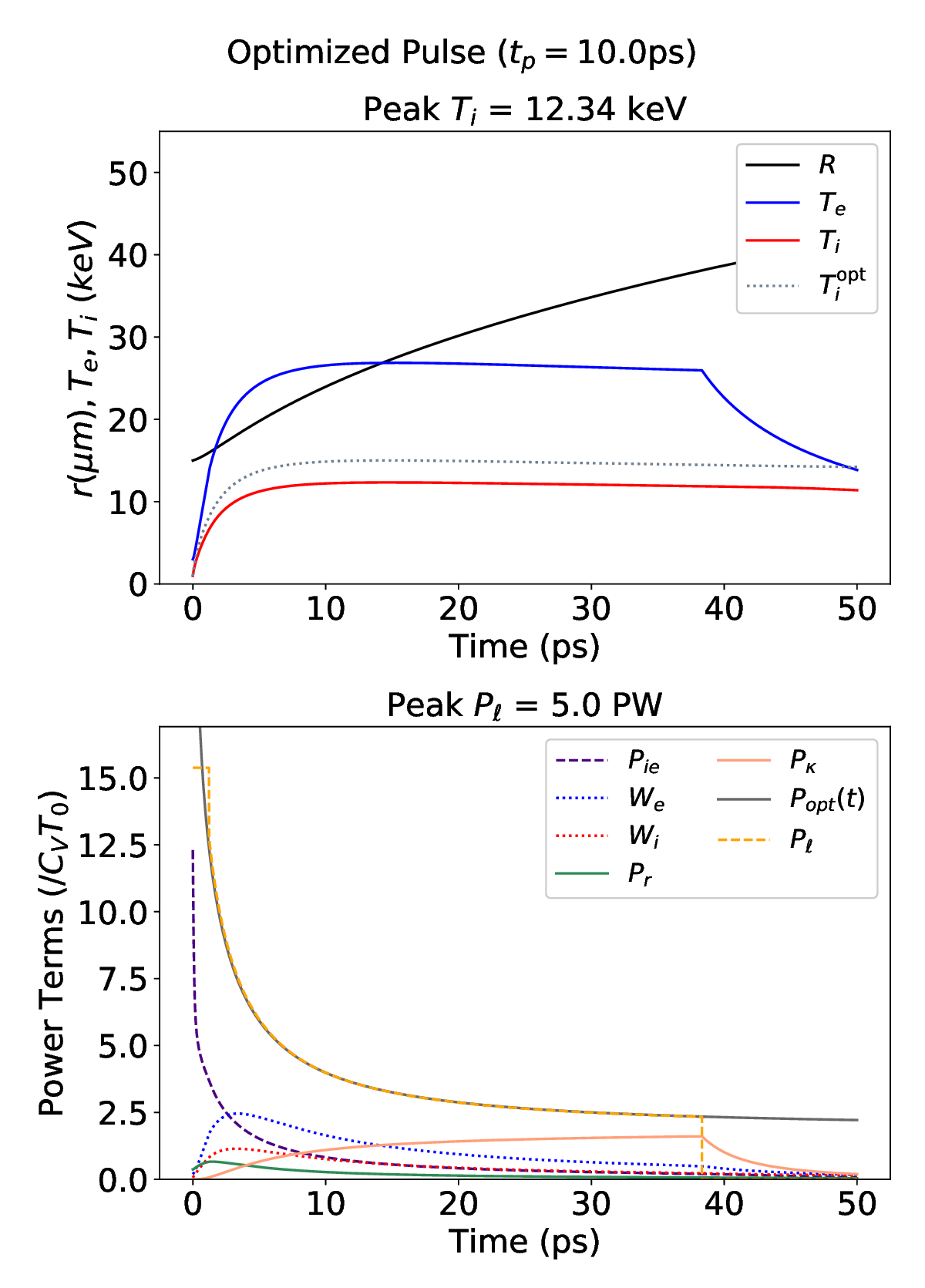}
\end{subfigure}
    \caption{Hotspot evolution for pulses with power $P_\mathrm{flat}(t)$ and $P_\mathrm{opt}(t)$. The top panels show $R$, $T_e$, and $T_i$ as functions of time. The bottom panels show power contributions to the temperature evolution as given in Eq.~\ref{eq_power_balance}. $P_\ell$ (dashed yellow line) represents the power deposition into electrons. $P_\mathrm{opt}(t)$ (grey line) represents the optimal pulse without power limits, given analytically in Eq.~\ref{eq_Pi_opt_t}. $T_i^\mathrm{opt}$ (dotted grey line) represents the resulting ion temperature.}
    \label{fig_pulse_time}
\end{figure}

As hotspot parameters, we took ${R_0 = 15 \mu\text{m}}$, ${T_0 = 1\text{keV}}$,\footnote{Although 1~keV is hotter than one would often want the initial isochoric compressed fuel to be, this initial condition for the hotspot helps to keep the model self-consistent by validating the assumptions of a strong shock, weak coupling, and nondegeneracy. This initial condition corresponds to starting electron temperature $T_e=3\text{keV}$, which is to be compared to the Fermi temperature $T_F\approx 763\text{eV}$. If the cold fuel starts at a few 100~eV, then our model can be taken to start tracking hotspot parameters after a very short period of heating where the hotspot goes from the cold fuel temperature to $T_0=1\text{~keV}$; the results are not meaningfully affected.} and ${\rho_0 = 400 \text{g/cm}^3}$. The pulse energy was held constant at ${50 \text{kJ}}$. Results are shown for a flat pulse and optimized pulse in Fig.~\ref{fig_pulse_time}. The top panel of each figure shows $R$, $T_e$, and $T_i$ as functions of time. 

The bottom panel of each figure shows the contributions of each term in Eq.~\ref{eq_power_balance}. The pulse power $P_\ell$ is shown by the dashed yellow line for the flat pulse and for the optimized pulse. For comparison $P_\mathrm{opt}$, the optimized pulse given in Eq.~\ref{eq_Pi_opt_t} if power were unlimited, is shown in grey.

\begin{figure}
    \centering
    \includegraphics[width=0.8\columnwidth]{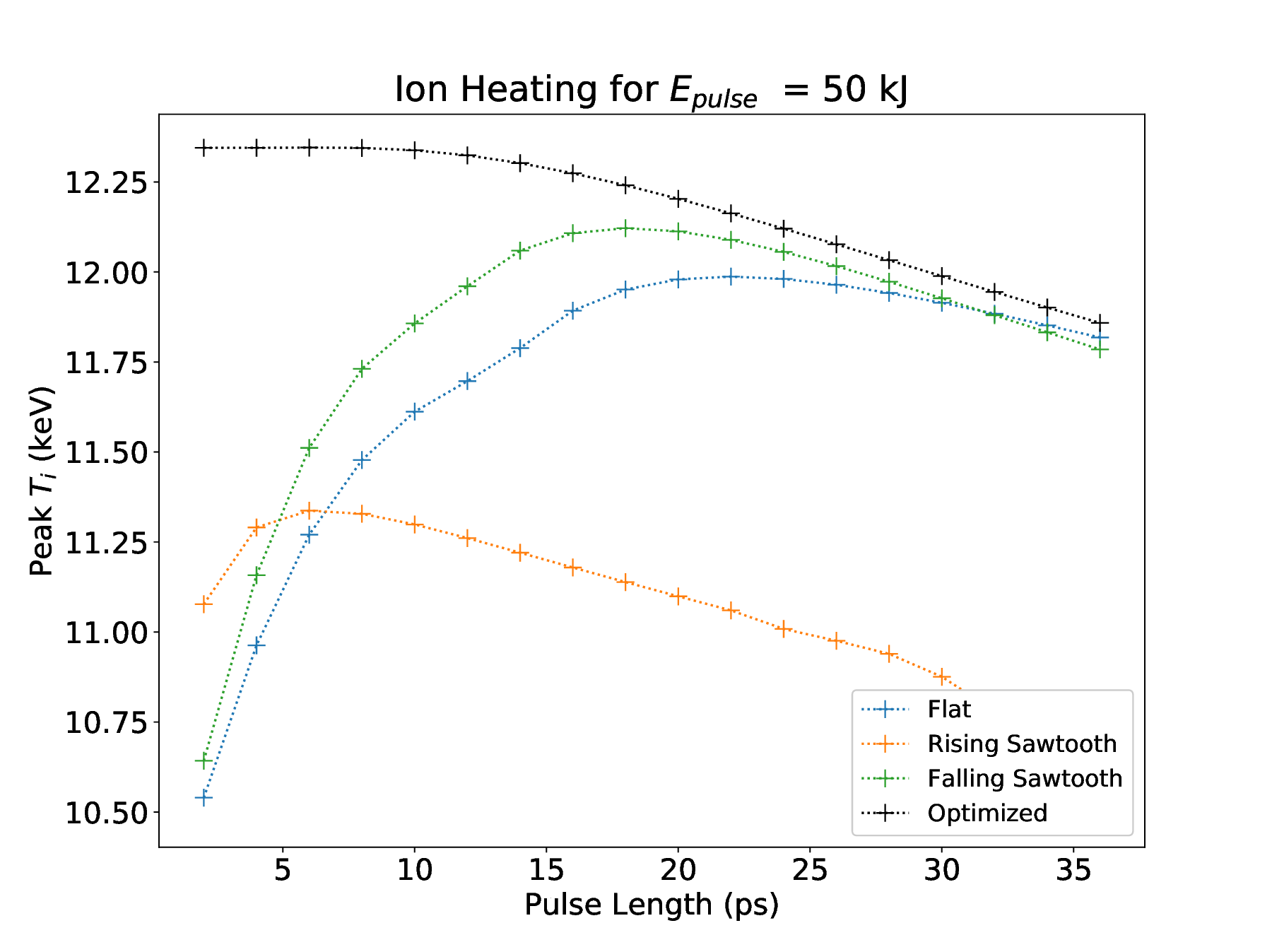}
    \caption{Maximum $T_i$ reached by ignitor pulse shapes as a function of $t_p$.}
    \label{fig_maxT_tp}
\end{figure}

We additionally scanned values of $t_p$, holding $E_p$ constant but letting $P_\mathrm{max}$ vary. The peak ion temperature attained by each pulse shape as a function of $t_p$ is shown in Fig.~\ref{fig_maxT_tp}. The three non-optimized pulse shapes exhibit maxima in peak $T_i$ with respect to $t_p$ because short, high-power pulses cause $T_e$ to overshoot its optimum, as seen in Fig.~\ref{fig_pulse_time}. By contrast, shortening $t_p$ for the optimized pulse never makes the performance worse; $P_\mathrm{max}$ is supplied only as long as advantageous, and power is reduced before $T_e$ overshoots its optimum. The performance does, however, saturate for small $t_p$ because $P_\mathrm{max}$ is no longer an important limitation. For long $t_p$, the peak ion temperatures of the optimized and flat pulses converge because the optimized pulse becomes power-limited, i.e. ${\Pi_\mathrm{opt} > \Pi_\mathrm{max}}$ for most of the pulse.

\section{Ignition Conditions}
\label{sec_ignition_conditions}

\subsection{Peak Ion Temperature}
\label{sec_peak_ti}

For a ignitor pulse of optimal shape, it is worthwhile to ask how hot the ions are able to become. The existence of a bound, which we show here, is a feature of the hotspot expansion combined with the temperature-dependent collision rate. As $\theta_i$ increases, $W_i$ also increases. Since the electrons must be hotter than the ions to heat them, $P_{ie}$ meanwhile decreases. Since $P_{ie}\sim x^{-3}$ and $W_i \sim x^{-5/2}$, expansion cannot reverse this trend. Eventually, ions will reach some critical $\hat \theta_i$ at which $\theta_i^\prime$ is not positive for any electron temperature, and so $\hat \theta_i$ represents the peak ion temperature that can be reached from given initial conditions.

To find $\hat \theta_i$, we observe that $\theta_i^\prime = 0$ at the peak radius $\hat x$, where ${\theta_i = \hat \theta_i}$ and ${\theta_e = c_0\hat\theta_i}$, and where $\hat x$ must also be found. In Appendix~\ref{app_peak_ti}, we derive $\hat \theta_i$ and the radius $\hat x$ at which it is reached. We define the dimensionless combination of constants appearing in these expressions as
\begin{equation}
\label{eq_kn_defn}
    \kn \doteq \frac{\nu c_1}{\sigma}
\end{equation}
and note that $\kn$ is proportional to the Knudsen number $\mathrm{Kn} \sim t_c/t_\nu$. We are also led to define a critical temperature $\mc T \doteq \kn^{1/2} T_0$ at which the collision rate and expansion rate become equal. The peak ion temperature $\hat T_i = T_0 \hat \theta_i$ is given from Eq.~\ref{eq_derivation_peak_Ti_final} as 
\begin{equation}
\label{eq_Ti_hat}
    \hat T_i = \frac{\mc T}{8^{1/14}\left[1 - (7/8)(T_0/\mc T)^2\right]^{1/14}} .
\end{equation}

If the hotspot starts far below the critical temperature, $T_0 \ll \mc T$, then an optimal pulse can raise $T_i$ to about $ 86\%$ of $\mc T$. If the hotspot starts at the critical temperature, $T_0 = \mc T$, then it begins at its peak, so even with an optimal pulse, $T_i$ will only decrease. Setting ion temperature $\Tign$ as a heuristic condition for ignition, Eq.~\ref{eq_Ti_hat} determines whether $\Tign$ can be reached by an optimal pulse from given initial conditions.

\subsection{Areal Density}
\label{sec_areal_density}

A self-heating hotspot requires both sufficient ion temperature and areal density $\rho R$. As a simple heuristic for ignition, we set the requirement ${\rho R > \adenig}$ as well as $T_i > \Tign$ where $\adenig = 0.4\text{~g/cm}^2$ and $\Tign = 10\text{~keV}$ \cite{Fernandez_Albright_Beg_Foord_Hegelich_Honrubia_Roth_Stephens_Yin_2014}. Since $\rho R = \rho_0 R_0/x^{2}$, expansion could bring the hotspot below the needed areal density even as ions are heated toward ignition temperature. Following from \S\ref{sec_params}, for a system-independent\footnote{More precisely, $k \sim \sqrt{\lambda}$, but we treat the Coulomb logarithm as a constant.} constant $k$, 
we have $\mc T = k \sqrt{\rho_0 R_0}$, which suggests determining whether a hotspot trajectory in ${(\rho R, T_i)}$ space will enter the ignition region by the following procedure, which applies when the ignitor pulse is optimal.

First, if $\hat T_i < \Tign$, the hotspot will never ignite within this model. If $\kn \gg 1$, which is often the case, then a necessary but not sufficient condition is approximately
\begin{equation}
\label{eq_aden_cond_approx}
\begin{split}
    \rho_0 R_0 &> \Tign^2 8^{1/7}/k^2 
    \\
    \rho_0 R_0 \uaden &> 3.08\times 10^{-9} \Tign \uT^2 .
\end{split}
\end{equation}

\begin{figure}
    \centering
    \includegraphics[width=0.8\columnwidth]{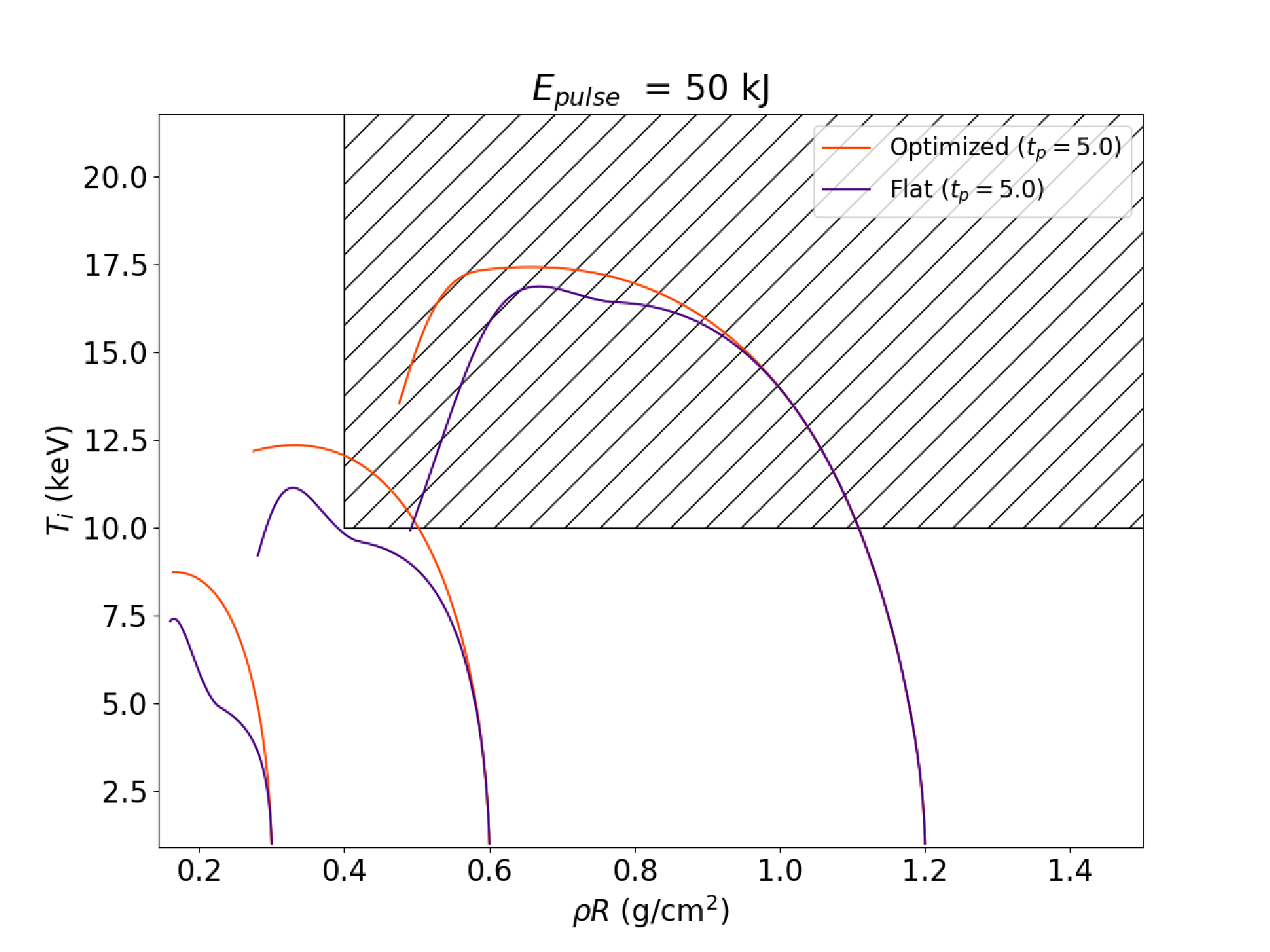}
    \caption{Hotspot trajectories in $(\rho R, T_i)$ space from $t=0$ to $t=25$ps for optimized and flat pulses. The shaded region is where $T_i>\Tign$ and $\rho R > \adenig$. For each areal density $\rho_0$ and $R_0$ as scaled so that $\rho_0 R_0^3$ is constant.}
    \label{fig_rhoR_T}
\end{figure}

If this condition is met, then define ${x_c \doteq \sqrt{\rho_0R_0 / \adenig}}$ as the normalized radius at which the hotspot becomes too diffuse to ignite. Then using Eq.~\ref{eq_theta_i_x} (and ${T_i = T_0\theta_i}$), calculate $T_i(x_c)$. If ${x_c>\hat x}$, ignition requires ${T_i(x_c) > \Tign}$; else ignition requires ${T_i(\hat x) > \Tign}$. For general pulse shapes, this condition requires numerical evaluation. Hotspot evolution is shown in Fig.~\ref{fig_rhoR_T} for several starting areal densities. The case of ${\rho R = 0.6\text{g/cm}^3}$ uses the same hotspot conditions as in \S\ref{sec_numerical_solutions}, and for the other curves $\rho_0$ and $R_0$ are scaled so that $C_V$ is constant. In all cases, the optimized pulse satisfies the ignition conditions better than the flat pulse.

\subsection{Pulse Energy}
\label{sec_pulse_energy}

The analysis above is performed for fixed $E_p$, but it is clear from Fig.~\ref{fig_rhoR_T} that the hotspot often reaches the ignition region before the entire pulse energy is spent. To improve fusion gain, it is valuable to reduce ignition energy. The pulse shape derived in this work (Eq.~\ref{eq_Pi_opt_ti_x}) is optimized for ion heating per increase in $R$ (equivalently, per decrease in $\rho R$), and not per unit pulse energy. Is $P_\mathrm{opt}(t)$ also energy efficient?

\begin{figure}
    \centering
    \includegraphics[width=0.8\columnwidth]{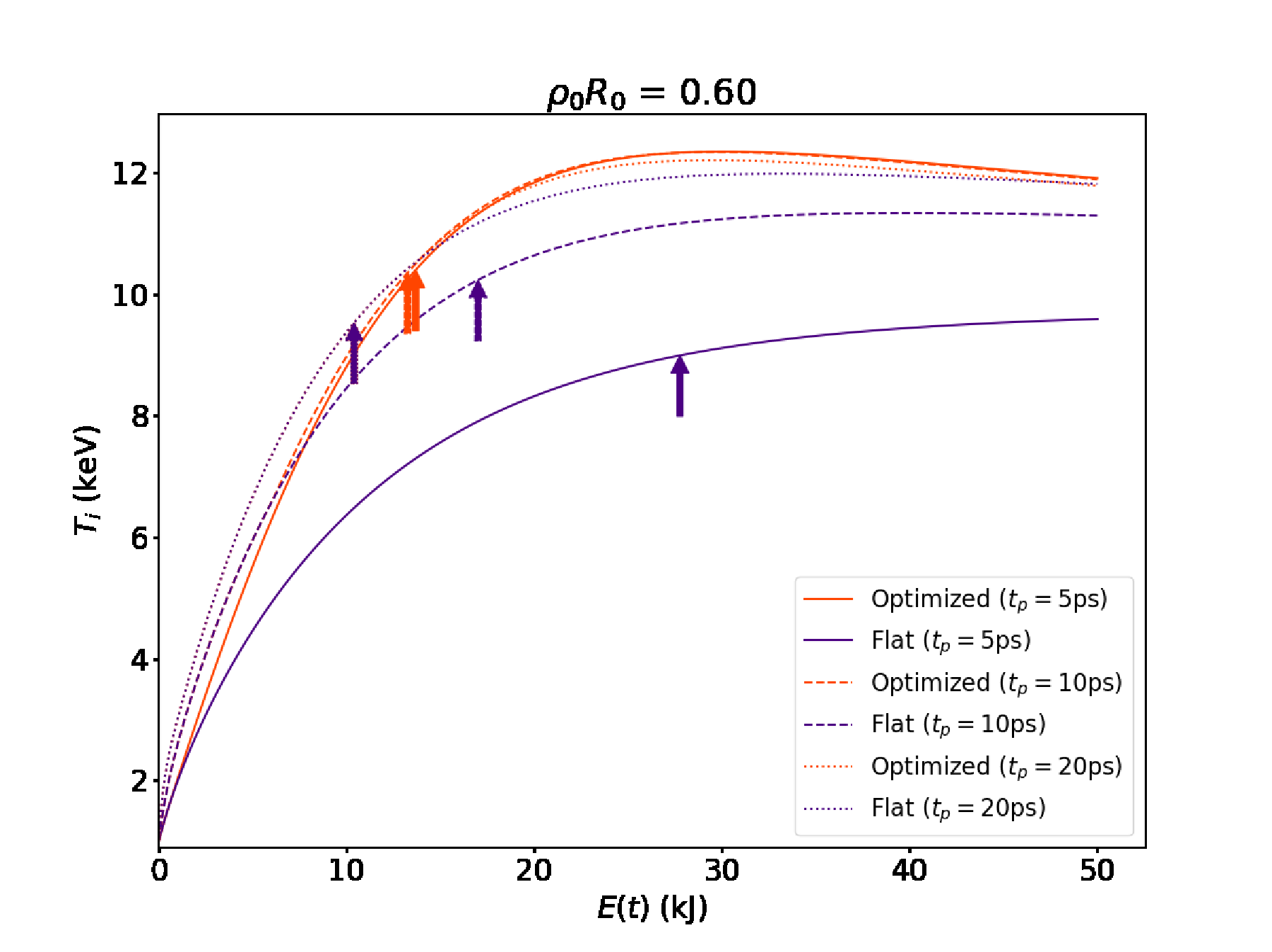}
    \caption{Ion temperature as a function of total energy $E(t)$ delivered by the ignitor at time $t$ for various pulse shapes. The arrows show the point at which areal density drops below $\rho R = \adenig$ due to hotspot expansion; by the ignition criteria used in this work, the hotspot must reach ignition temperature $T_i=\Tign$ before this point.}
    \label{fig_T_Et}
\end{figure}

In Fig.~\ref{fig_T_Et}, $T_i(t)$ is shown during a pulse as a function of $E(t)$, the total pulse energy that has been delivered at time $t$. The pulses had total energy $E_p = 50\text{kJ}$ and characteristic times $t_p = 5\text{~ps}, 10\text{~ps}, 20\text{~ps}$. The arrows show where each curve crosses ${x = x_c}$; ignition needs to happen before this point, as discussed in \S~\ref{sec_areal_density}.

By the ignition criteria in \S~\ref{sec_areal_density} and for $t_p = 10\text{~ps}$, the optimized pulse requires $23\%$ less energy to reach $\Tign$ compared to the flat pulse. This margin varies significantly for different hotspot and ignitor parameters.

The profiles for optimized and flat pulses initially have the same shape because both pulses are power-limited. For the longer $t_p$ (lower power) case, this continues until $T_i \approx 7\text{keV}$, whereas for the shorter $t_p$ case, the profiles quickly diverge. In contrast to Fig.~\ref{fig_maxT_tp} where shorter $t_p$ was strictly better for the optimized pulse, we see here that the 10~ps pulse is slightly more energy efficient than the 5~ps for $E \lesssim 20\text{kJ}$ and, for $E \lesssim 7\text{kJ}$, even the 10~ps flat pulse is more energy efficient than the 5~ps optimized pulse. With higher maximum power, $P_\mathrm{opt}$ is free to drive the electrons to $T_e = c_0 T_i$, in which case only $\sim 31\%$ of hotspot energy is in the ions. Although this energy partition leads to the fastest ion heating (per radial increment), it means that a larger share of pulse energy has gone into electrons.

We note, however, that gains in energy efficiency by reducing $T_e$ come at the cost of decreased $\rho R$. This is visible in Fig.~\ref{fig_T_Et} where the 20~ps pulse, albeit the most energy efficient for $E \lesssim 14\text{kJ}$, is the first to fall below $\adenig$. Additionally, under more realistic ignition criteria, $\Tign$ should depend inversely on $\rho R$ \cite{Atzeni_Meyer-Ter-Vehn_2004,Tabak_Hinkel_Atzeni_Campbell_Tanaka_2006}, as opposed to the independent inequalities used here as a heuristic. The majority of the energy for FI is spent by the initial drivers to compress fuel to required $\rho_0 R_0$, rather than by the ignitor. Therefore, it is likely that the full process energy efficiency is generally higher for pulse shapes that optimize $T_i(\rho R)$ as opposed to $T_i(E)$, but the details of this tradeoff require further study.

\section{Discussion}
\label{sec_discussion}

The reduced model presented in this work captures the key features of ion heating in a uniformly expanding two-temperature plasma. We have aimed to isolate the important physics behind initial phases of hotspot heating for FI, which can aid in designing methods to improve heating efficiency. Here, we identify three directions in which the model could be expanded to make better quantitative prescriptions for ignitor design.

First, the fidelity of the model could be increased. Several possible extensions can be made while preserving the ability to optimize analytically, but these are beyond the scope of the present work. Burning plasma physics could be incorporated, with alpha particles depositing energy in hotspot electrons, hotspot ions, and surrounding cold fuel \cite{Christopherson_Hurricane_Weber_Kritcher_Nora_Salmonson_Tran_Milovich_Maclaren_Hinkel_et}. The expansion model, while based on common estimates in existing work \cite{Atzeni_Meyer-Ter-Vehn_2004,Piriz_Sanchez_1998,Tabak_Hinkel_Atzeni_Campbell_Tanaka_2006}, could be improved upon; ideally, the uniformity assumption would be relaxed and the model would involve solving the fluid equations in at least the radial dimension. Power deposition could also be modeled in a more realistic, nonuniform way, including a generally nonspherical distribution. Dependence of power deposition on the evolving density and temperature of the hotspot can be important \cite{Piriz_Sanchez_1998}.

Second, the optimization criteria could be adjusted to maximize total gain (fusion power divided by combined driver and ignitor energy). As discussed in \S\ref{sec_pulse_energy}, peak $T_i$ is a reasonable proxy but does not tell the full story. The efficiency of the ignitor, which may vary depending on the peak power it is called upon to provide, could also be included.

Third, most ignitor designs use an ultrafast laser to produce fast particles near the outer edge of the fuel capsule, which propagate into the hotspot. In this work, we take $P_\ell$ to be the power that is ultimately deposited in hotspot electrons. Using a desired $P_\ell$ profile to determine a profile for the original laser will pose a difficult inverse problem, but its results could affect calculations of the ignitor efficiency. For example, experimental studies of proton acceleration from foils using pulsed lasers have shown that the properties of accelerated protons depend on pulse duration as the foil plasma evolves during the pulse, with longer pulses achieving higher efficiency and maximum proton energy \cite{Fuchs_Antici_d’Humières_Lefebvre_Borghesi_Brambrink_Cecchetti_Kaluza_Malka_Manclossi_et,Yogo_Mima_Iwata_Tosaki_Morace_Arikawa_Fujioka_Johzaki_Sentoku_Nishimura_et}. These works, as well as simulations \cite{Kim_Kemp_Wilks_Kalantar_Kerr_Mariscal_Beg_McGuffey_Ma_2018}, have shown that the energy of accelerated protons rises over the course of a multi-picosecond pulse. The corresponding idealized `rising sawtooth' pulse shape (cf. Eq.~\ref{eq_pulse_forms} and Fig.~\ref{fig_maxT_tp}) is the least effective of the pulse shapes considered in this work for heating ions. The goals of maximizing ion heating and ignitor efficiency appear therefore to be in conflict; this adds an important new dimension to the optimization, which is beyond the scope of this work.

Some ignitor designs do not only heat electrons. Heavy ion beams for example \cite{Fernandez_Albright_Beg_Foord_Hegelich_Honrubia_Roth_Stephens_Yin_2014,Hegelich_Jung_Albright_Fernandez_Gautier_Huang_Kwan_Letzring_Palaniyappan_Shah_et, Regan_Schlegel_Tikhonchuk_Honrubia_Feugeas_Nicolai_2011} have the advantage of depositing a significant fraction of their energy directly into ions. Our demonstration that ion heating by electrons is severely limited under certain relevant conditions may influence the desirability of these alternative schemes.

\section{Conclusions}
\label{sec_conclusions}

The commonly assumed flat pulse shape for fast ignition has been shown to drive suboptimal ion heating in an expanding hotspot. An optimized pulse shape has been derived analytically in terms of hotspot parameters (Eq.~\ref{eq_Pi_opt_ti_x}) and presented as an explicit function of time (Eq.~\ref{eq_Pi_opt_t}) in the case where unlimited ignitor power is available. A bound has been derived on the maximum ion temperature that can be reached by electron heating only (Eq.~\ref{eq_Ti_hat}). 

Non-optimized pulse shapes become less effective at higher power, but the optimized pulse can take advantage of greater available power to increase the rate of ion heating with respect to hotspot expansion. The pulse performance is evaluated against heuristic criteria for ignition and it is shown that optimizing pulse shape can push a hotspot's trajectory into the ignition region of $(\rho R, T_i)$ space. Gains in efficiency in the sense of ion temperature per unit ignitor power are modest for long pulses but substantial for short pulses, reaching $23\%$ in a representative case.

Analysis has been done on a reduced model in which a sphere of two-temperature plasma expands uniformly. This captures the essential physics of the early stages of hotspot heating and arrives at interesting, nonintuitive results for the dependence on pulse parameters. Formulating an analytical description allowed us to reduce the complex task of designing an ignitor pulse shape to a straightforward optimization problem. These analytical results are useful in disentangling the competing physical effects at work in hotspot heating, and for computational studies that further refine the optimal pulse shape using higher-fidelity models, our results in Eq.~\ref{eq_Pi_opt_ti_x} offer a valuable starting point for reducing the size of the parameter space to be explored.

Further work could extend our analysis into the burning plasma regime, but the results here for non-fusing plasma may have immediate relevance for downscaled FI experiments. A wide variety of other experimental platforms involve heating electrons in an expanding plasma \cite{Rochau_Bailey_Falcon_Loisel_Nagayama_Mancini_Hall_Winget_Montgomery_Liedahl_2014,Sinars_Sweeney_Alexander_Ampleford_Ao_Apruzese_Aragon_Armstrong_Austin_Awe_et,Kraus_Gao_Fox_Hill_Bitter_Efthimion_Moreau_Hollinger_Wang_Song_et} and could benefit from these results.

This work was supported by DOE Grant 83228–10966 [Prime No. DOE (NNSA) DE- NA0003764].

\newpage

\appendix 

\section{Optimal Pulse in Time}
\label{app_temporal_shape}

Using Eq.~\ref{eq_theta_i_x} for $\theta_i(x)$ to solve for the evolution of $x$ in Eq.~\ref{eq_x_dot}, we have
\begin{equation}
    \dot x = \frac{\sigma}{2}x^{-5/2} \sqrt{c_0 + 1}\left(1 + \frac{8}{7}\kn (x^{7/2} - 1)\right)^{1/4}.
\end{equation}

This equation is separable and we can solve it, noting that the initial conditions are $t=0, x=1$ by definition, to obtain
\begin{equation}
    \left(1 + \frac{8}{7}\kn (x^{7/2} - 1)\right)^{3/4} - 1 = \frac{3}{2}\kn \sigma \sqrt{c_0 + 1} t .
\end{equation}

We then solve for $x$ and find
\begin{equation}
\label{eq_x_t}
    x(t) = \left[1 + \frac{7}{8\kn}\Big(w(t)^4 - 1\Big)\right]^{2/7}
\end{equation}
where
\begin{equation}
\label{eq_w_t}
    w(t) \doteq \frac{3}{2} \nu c_1\sqrt{c_0 + 1}t + 1
\end{equation}
and we note that
\begin{equation}
    \theta_i(t) = \frac{w(t)^2}{x(t)^2} .
\end{equation}

\begin{widetext}
Finally, using the prescription in Eq.~\ref{eq_Pi_opt_ti_x} and writing $\Pi_\mathrm{opt}(t)$ in terms of $w(t)$ (Eq.~\ref{eq_w_t}) for brevity, we arrive at
\begin{equation}
\label{eq_Pi_opt_t}
\begin{split}
    \Pi_\ell(t) =&  \frac{\nu c_1\sqrt{c_0+1}}{w(t)\left[1 + \frac{7}{8\kn}\Big(w(t)^4 - 1\Big)\right]^{4/7}} + \frac{\beta c_0^{1/2}w(t)}{\left[1 + \frac{7}{8\kn}\Big(w(t)^4 - 1\Big)\right]^{8/7}} + \frac{\kappa c_0^{7/2} w(t)^7}{\left[1 + \frac{7}{8\kn}\Big(w(t)^4 - 1\Big)\right]^{12/7}} .
\end{split}
\end{equation}
\end{widetext}

\section{Location of Temperature Peak}
\label{app_peak_ti}

We are interested in the point $(\hat x, \hat \theta_i)$ beyond which ion temperature can no longer increase. This means that $\theta_i^\prime(\hat x) = 0$ when electron temperature takes its optimal value $\theta_e = c_0\theta_i$ (and $\theta_i^\prime < 0$ for nonoptimal $\theta_e$). From Eq.~\ref{eq_theta_i_prime} we have
\begin{equation}
    0 = -2\frac{\hat\theta_i}{\hat x} + \frac{2\nu}{\sigma} \frac{c_0 - 1}{c_0^{3/2} (c_0+c_1)^{1/2}\hat x^{3/2}\hat \theta_i}
\end{equation}
which simplifies to
\begin{equation}
    \hat\theta_i^2 = \kn \hat x^{-1/2} .
\end{equation}

Now using Eq.~\ref{eq_theta_i_x}, we have
\begin{equation}
    \hat\theta_i = \frac{\hat\theta_i^8}{\kn^4}\sqrt{1 + \frac{8}{7}\kn (\kn^7\hat\theta_i^{-14} - 1)}.
\end{equation}

This simplifies to
\begin{equation}
\label{eq_derivation_peak_Ti_final}
    \frac{\kn}{7}\kn^7\hat\theta_i^{-14} = \frac{8}{7}\kn - 1
\end{equation}
from which one can readily obtain Eq.~\ref{eq_Ti_hat}.



\bibliography{pulseshaping}

\end{document}